\documentclass[journal,print,onecolumn]{ieeecolor}
\usepackage{generic}
\usepackage{amsmath,amssymb,amsfonts}
\usepackage{algorithmic}
\usepackage{graphicx}
\usepackage{algorithm,algorithmic}
\usepackage{textcomp}
\usepackage{multirow}
\usepackage{float}

\usepackage{caption}
\usepackage[style=authoryear]{biblatex}
\usepackage{hyperref}
\hypersetup{hidelinks=true}

\begin{document}

\title{Identifying Body Composition Measures That Correlate with Self-Compassion and Social Support
}
\def\BibTeX{{\rm B\kern-.05em{\sc i\kern-.025em b}\kern-.08em
 T\kern-.1667em\lower.7ex\hbox{E}\kern-.125emX}}
\markboth{\hskip25pc}
{Enerson Poon \MakeLowercase{\textit{et al.}}: Identifying Body Composition Measures That Correlate with Self-Compassion and Social Support.}

\author{Enerson Poon \textsuperscript{1}, Mikaela Irene Fudolig, Johanna E. Hidalgo, Bryn C. Loftness, Kathryn Stanton, Connie L. Tompkins, Laura S. P. Bloomfield, Matthew Price, Peter Sheridan Dodds, Christopher M. Danforth, Nick Cheney
 \\
 \textit{University of Vermont, Burlington, USA} \\}

\maketitle
\footnotetext[1]{Corresponding Author: epoon@uvm.edu}

\date{10/11/2025}

\setcounter{secnumdepth}{0}

\maketitle

\begin{abstract}
This study explores the relationship between body composition metrics, self-compassion, and social support among college students. Using seasonal body composition data from the InBody770 system and psychometric measures from the Lived Experiences Measured Using Rings Study (LEMURS) (n=156; freshmen=66, sophomores=90), Canonical Correlation Analysis (CCA) reveals body composition metrics exhibit moderate correlation with self-compassion and social support. Certain physiological and psychological features showed strong and consistent relationships with well-being across the academic year. Trunk and leg impedance stood out as key physiological indicators, while \textit{mindfulness}, \textit{over-identification}, \textit{affectionate support}, and \textit{tangible support} emerged as recurring psychological and social correlates. This demonstrates that body composition metrics can serve as valuable biomarkers for indicating self-perceived psychosocial well-being, offering insights for future research on scalable mental health modeling and intervention strategies.
\end{abstract}


%
\IEEEpeerreviewmaketitle

\section*{Introduction}

Studies show that psychological states shape everyday behaviors, and that these behaviors, such as physical activity (\cite{mahindru_role_nodate}), dietary habits (\cite{heidari_influence_2023}), sleep (\cite{Hyndych2025}), and substance use (\cite{corace_substance_2022}), in turn influence physical health. These effects extend beyond emotional well-being and are reflected in measurable biological processes throughout the body. For example, there are strong associations between depression and cardiovascular disease (\cite{kwapong_association_2023}).

Psychosocial factors such as social support and self-compassion appear to regulate both mental health and acute/chronic stress within the body (\cite{chu_physiology_2026,ross_distress_2025}). Because stress influences multiple physiological systems, its effects persist over time and contribute to long-term health consequences (\cite{Shchaslyvyi2024}). For example, chronic stress has been associated with lasting physical outcomes such as cardiovascular morbidity (\cite{satyjeet_psychological_nodate}).

Body composition is an important downstream domain through which the regulation of the stress pathway exerts measurable effects. Because acute and chronic stress responses influence cortisol activity (\cite{knezevic_role_2023}), immune signaling (\cite{alotiby_immunology_2024}), metabolic processes (\cite{sic_neurobiological_2024}), and cardiovascular function (\cite{satyjeet_psychological_nodate}), they also shape patterns of fat storage, muscle maintenance (\cite{stefanaki_chronic_2023}), and water retention (\cite{flomenbaum_morphologic_2022}).

Among the psychosocial factors that may shape these stress-mediated changes in body composition, self-compassion and social support are particularly relevant for college students. These factors have been shown to affect university students’ ability to confront and recover from adversity, contributing to a greater sense of meaning in life (\cite{chan_impact_2022}). In turn, having strong social networks and self-compassion boosts positive behaviors (\cite{wang_association_2024}) and promotes mental health among freshmen (\cite{wei_first-year_2022}). 

Previous studies have shown there is a link between self-compassion, well-being, and resilience (\cite{neely_self-kindness_2009, lavin_role_2020}) and that these linkages could be related to body mass index (BMI) and body perception (\cite{nightingale_self-compassion_2023}). However, research examining the relationship between BMI and self-compassion is limited, and recent studies collectively diverge in their use of BMI as a biomarker (\cite{pray_history_2023}). 

On a societal scale, labeling people at a young age as “fat” or “obese” can lead to decreased emotional well-being and increased discrimination. From a medical perspective, BMI is a poor representation of the total body fat percentage and lacks precision in estimating fat mass across regions (\cite{potter_defining_2024}). Recent research suggests that mortality risk remains relatively low across a broad range of BMIs when comparing individuals within similar age groups, typically among adults in middle age (\cite{nuttall_body_2015}). However, BMI alone does not adequately capture mortality risk, as numerous factors—including comorbidities, lifestyle, gender, ethnicity, genetics, duration within BMI categories, and fat distribution—significantly influence health outcomes (\cite{nuttall_body_2015}). Mortality risk here refers to relative differences over specific time frames rather than the absolute certainty of death, highlighting that BMI is not a definitive predictor without context. These confounders must be carefully considered before using BMI data as the sole basis for public health policies targeting morbidity and mortality. Therefore, using BMI as a definitive measure of health or mortality risk can be misleading (\cite{visaria_body_2023}). 

Rather than relying on BMI, recent research has shown that body composition metrics relate to psychosocial measures and academic success, emphasizing the importance of incorporating these more nuanced measures in evaluating overall well-being (\cite{redondo-florez_body_2021}).

We argue that body composition measures could be directly linked to psychosocial factors due to the connection between mental and physical states. For example, studies indicate that self-compassion is linked to a range of health-promoting behaviors, including lower smoking rates, healthier eating and exercise habits, greater likelihood of seeking medical care, higher physical activity levels, safer sexual practices, and reduced bedtime procrastination (\cite{biber_effect_2019, Bistricky2025, wang_influence_2014}) . Additionally, identifying new indicators of psychosocial factors using body composition metrics could help early identify mental health issues in colleges and universities, where scalable, cost-effective solutions are unavailable or challenging to implement (\cite{ebert_barriers_2019,peterson_barriers_2024}). 

Only 35.2\% of students report having received psychological or mental health services (United States) in the last 12 months (at the time of this study) (\cite{ncha_survey_summary_2024}). Colleges and universities within the United States need to identify more socially accepted, affordable, and non-invasive ways to provide mental health services. Finding other ways to measure and quantify mental health could be the key to expanding the reach of mental health services. 

Furthermore, only a small portion of colleges/universities in the United States offer mental health services(\cite{aluri_prevalence_2025}) and only ~19\% of college students receive psychotherapy (\cite{rackoff_psychotherapy_2025}). InBody770 machines are cost-effective, non-invasive, and could be incorporated into a standard primary care check-up. This research advocates for making these tools more accessible for surveying mental health, if they are not already available.

Finally, unlike traditional psychotherapy, InBody770 machines (and similar body composition analyzers) are quick, easy to interpret, and user-friendly. A decreasing trend in diagnostic mental health assessments and mental health services in public schools could indicate (\cite{noauthor_coe_2024}) that body composition analyzers are a promising avenue for identifying mental health markers. 

To date, there is limited, well-established evidence demonstrating the direct relationships between self-compassion, social support, and body composition within college students and at the depth of our study. While research on self-compassion and body composition has primarily focused on meditation or body image (\cite{tolonen_compassion_2025, nightingale_self-compassion_2023-bodyimage, james_mechanisms_2023}), the relationship between social support and body composition is more well established (\cite{wang_influence_2014, tymoszuk_social_2019}). Our study argues that body composition analyzers could serve as tools for identifying markers of declining mental health and supporting providers in advocating for appropriate mental health services.

Utilizing a dataset of 156 first- and second-year college students from the Lived Experiences Measured Using Rings Study (LEMURS) (\cite{price_large_2023, fudolig_two_2024, bloomfield_predicting_2024, bloomfield_anxietypredictors_2024}), we examine seasonal change in body composition measures and find novel biomarkers for self-compassion and social support. Instead of relying solely on BMI, we use metrics from an InBody770 (\cite{mclester_reliability_2020}), a body composition analyzer that provides the breakdown of body mass into various components, including fat, muscle, and water. We explore the relationship between these biometrics and self-reported psychosocial measures of self-compassion and social support.

To assess whether correlations between body composition, self-compassion, and social support should be examined across subcohorts (e.g., freshmen vs. sophomores or seasonal groups), we investigated the influence of the college experience on self-compassion and social support. Specifically, we explore whether sophomores—having already navigated their first year—exhibit different patterns of self-compassion and perceived social support compared to freshmen, who are experiencing this transition for the first time (\cite{worsley_bridging_2021,mclean_perceived_2023}).  We hypothesize that prior exposure to the college environment shapes psychological and social well-being, producing measurable differences across groups. Additionally, we hypothesize that meaningful group-level correlations exist between body composition and self-reported psychosocial measures of self-compassion and perceived social support.

\section*{Background}

Self-compassion has been defined as showing kindness and understanding toward yourself when faced with personal flaws, errors, failures, and challenging life circumstances (\cite{raes_construction_2011, neff_self-compassion_2023}). Research has increasingly highlighted the role of self-compassion in enhancing psychological well-being among college students. \cite{neely_self-kindness_2009} investigated the relationship between self-compassion, goal regulation, and support, finding that higher levels of self-compassion were associated with better emotional coping skills and overall well-being. Building on this, \cite{smeets_meeting_2014} conducted a study with female college students, demonstrating that a brief three-week self-compassion intervention significantly improved resilience and well-being. Further exploring this concept, \cite{lavin_role_2020} examined the impact of self-compassion on perceived social support among undergraduates, revealing that students with higher self-compassion reported stronger feelings of support from friends. These studies suggest that fostering self-compassion is a valuable strategy to improve mental health and social connectedness in the college student population.

 Our  study examines the connections between body composition metrics, social support, and self-compassion to better understand how these psychosocial factors are reflected in the body. Identifying new indicators of psychosocial factors through body composition metrics could aid in the early detection of mental health issues within colleges and universities, especially in settings where scalable, cost-effective solutions are limited or difficult to implement.

\section*{Methods}

Between September 2023 and April 2024, the Lived Experiences Measured Using Rings Study (\cite{price_large_2023}) collected InBody770 and main assessment data (data comprising all surveys and psychological assessments) once per season (September, January, and April). Seasonal (fall, winter, and spring) body composition was measured with InBody770. From this point forward, these measurement periods will be referred to as fall, winter, and spring.

Data collected under IRB\# 00000485, approved on 7/11/2023, via the University of Vermont and State Agricultural College: FWA 00000723 and The UVM Medical Center: FWA 00000727.

\subsection{Measures}

This study uses the Self-Compassion Scale-Short Form (SCS-SF)(\cite{neff_development_2021, neff_self-compassion_2023, neff_self-compassion_2010}) to evaluate self-compassion and calculate several subscales that measure different modalities(\cite{raes_construction_2011}) . In this instrument, \textit{self-kindness} refers to the practice of responding to personal flaws and shortcomings with care and understanding, rather than harsh self-criticism (\cite{raes_construction_2011, neff_self-compassion_2023}). It involves offering warmth and unconditional acceptance, even while acknowledging areas for growth. In the face of stress, a self-compassionate approach involves pausing to provide comfort rather than trying to immediately resolve the issue.

The component of \textit{common humanity} in self-compassion emphasizes the recognition of shared human imperfection (\cite{raes_construction_2011, neff_self-compassion_2023}). This concept is important in understanding how individuals relate to their struggles in the context of broader human experiences, which may influence emotional resilience during periods of stress or transition. Similarly, \textit{mindfulness} in self-compassion involves maintaining awareness of one’s painful emotions in a balanced manner, preventing \textit{over-identification} (recognizing someone or something as having a particular problem or characteristic that they do not actually have) with distress (\cite{raes_construction_2011, neff_self-compassion_2023, Lindsay2017}) . This is particularly relevant to our study, as it may help explain how individuals navigate emotional challenges, such as homesickness or academic pressure, during the college transition. \textit{Isolation} refers to the belief or feeling that one’s suffering, failures, or flaws are unique and separate—that others are happy or functioning well while one is alone in their struggles. By examining these concepts, we can gain insight into how self-compassion may impact emotional resilience among students and their ability to manage stress.

The concept of social support, as defined by \cite{sherbourne_mos_1991}, encompasses both the perception and reality of being cared for, valued, and included by others. It highlights the multidimensional nature of support, including emotional, informational, and practical assistance during times of need. Additionally, the definition emphasizes the diverse sources of support, such as family, friends, and community, underscoring their role in helping individuals cope with challenges and stressors.

The Medical Outcomes Social Support Survey (MOSS) includes several subscales that measure different aspects of social support (\cite{sherbourne_mos_1991}). The \textit{emotional support} subscale assesses the availability of individuals to confide in, listen to personal concerns, provide advice, and offer informational support when needed. The \textit{tangible support} subscale evaluates the availability of practical help with daily tasks, such as assistance with chores, meal preparation, or transportation. The \textit{affectionate support} subscale measures the availability of emotional warmth, including love, affection, and physical comfort, such as hugs, and the extent to which individuals feel wanted. The \textit{positive social interaction} subscale assesses the availability of individuals with whom one can engage in enjoyable activities or social interactions for relaxation or fun. Finally, the total score (\textit{overall social support}) encompasses all 19 items from the survey and measures the general availability of social support across these various domains.

Our study utilized data collected on the InBody770 body composition analyzer (InBodyUSA, Cerritos, CA).
The InBody770 is a high-precision body composition analyzer that uses bioelectrical impedance analysis (BIA) to measure and differentiate between key components of the body: fat, muscle, water, and more. It sends a low-level electrical current through the body at multiple frequencies to determine the resistance and reactance of various tissues. Unlike simpler BIA machines, the InBody770 utilizes segmental analysis, which means it independently measures the composition of each arm, leg, and trunk. It also separates intracellular water from extracellular water, offering detailed insights into inflammation, hydration status, and overall cellular health. With no empirical estimations based on age, gender, or ethnicity, the machine provides direct, reproducible data that is often used in clinical, athletic, and research settings to monitor health, track performance, and guide intervention strategies.

While traditional body composition refers to metrics such as fat mass, lean mass, and total body water—typically derived from bioelectrical impedance analysis (BIA)—our approach expands this definition to include complementary physiological indicators. Variables such as impedance, heart rate (HR), systolic blood pressure (SBP), and diastolic blood pressure (DBP) are not direct measures of body composition, nor are they core outputs of the InBody770. However, impedance values reflect the electrical properties of tissues and fluids, providing a window into physiological variation not captured by standard body composition metrics. Similarly, HR (a proxy for autonomic balance), SBP (a marker of vascular load), and DBP (a reflection of peripheral resistance)—though unrelated to the impedance-based BIA—offer additional physiological context. In our analysis, including these variables significantly enhanced the strength and interpretability of canonical correlations between physiological and psychosocial factors. To ensure conceptual clarity, we explicitly acknowledge these as physiologically relevant but non-traditional body composition variables that enrich, rather than confound, our modeling framework. We will refer to these physiological indicators throughout the paper as body composition metrics for simplicity and consistency, recognizing that their shared role in capturing internal physiological states makes them functionally relevant in the context of our multivariate analysis.

\subsection{Data Processing}

Survey assessment data and InBody770 data were preprocessed and a k-nearest neighbors imputer from sklearn (\cite{pedregosa_scikit-learn_2018}) was utilized to impute any missing data using information from similar instances. Machine learning and statistical studies commonly use this technique to address small amounts of missing data, and it is considered a standard approach for data imputation in such cases. The prevalence of missing data is low, with only 3.3\% of the main assessment data and 0.25\% of the InBody770 data needing imputation.

\begin{figure*}
 \centering
 \includegraphics[width=\linewidth]{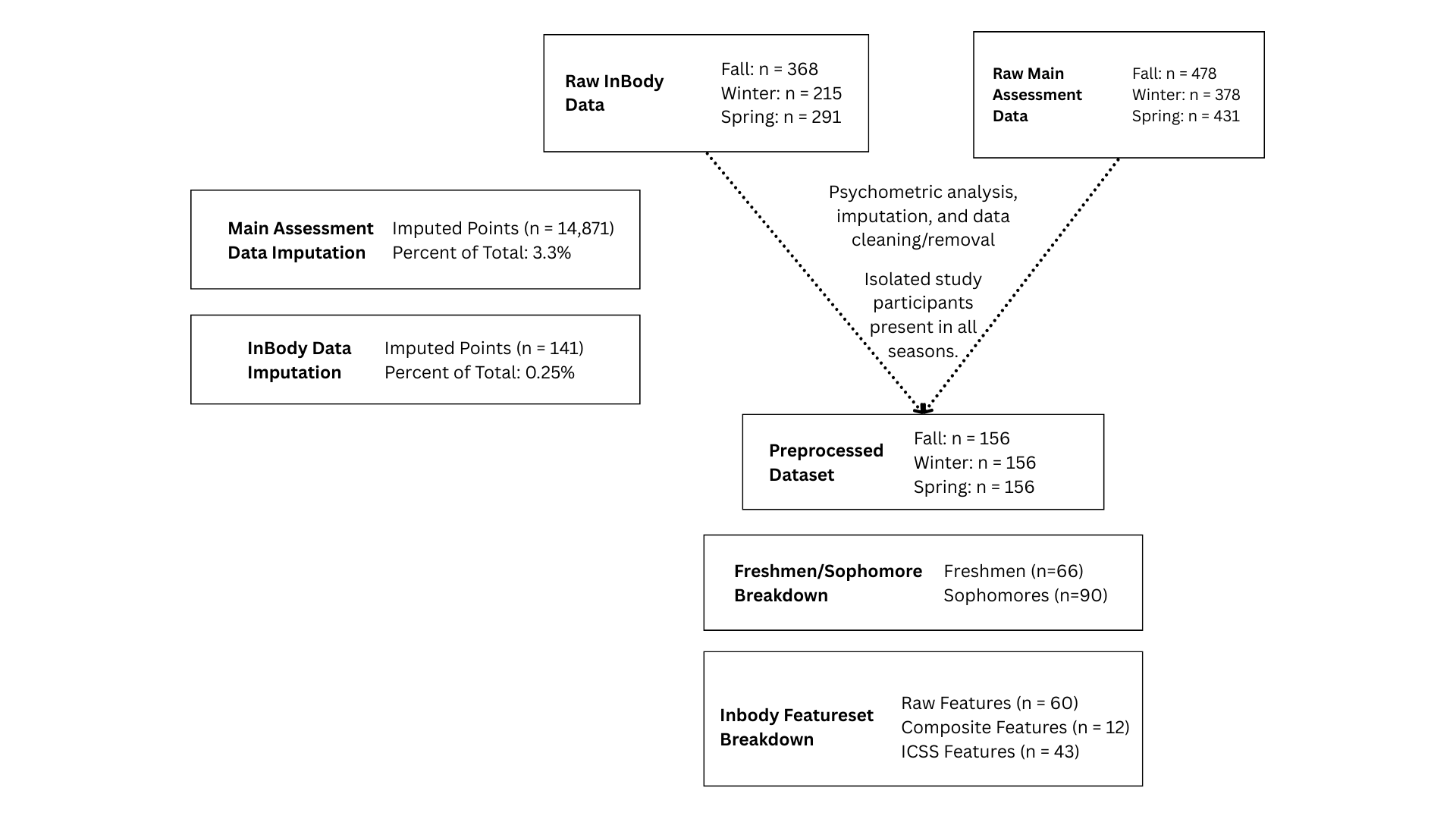}
 \caption{Diagram illustrating participant attrition and data imputation throughout the preprocessing stage of the study. Note the level of attrition through the preprocessing stage; a total of 156 students (66 freshmen and 90 sophomores) overlapped in all three seasonal sessions.}
 \label{fig:preprocessing}
\end{figure*}

Students who completed both surveys and InBody770 measurements across all three seasons were included in this study (N=156) (\ref{fig:preprocessing}). Psychometric analysis (PA) (\cite{raes_construction_2011, sherbourne_mos_1991}) was conducted on Self-Compassion and MOSS survey measures to create PA composite measures utilized throughout our study.

\begin{figure}[h]
 \centering
 \includegraphics[width=0.49\linewidth]{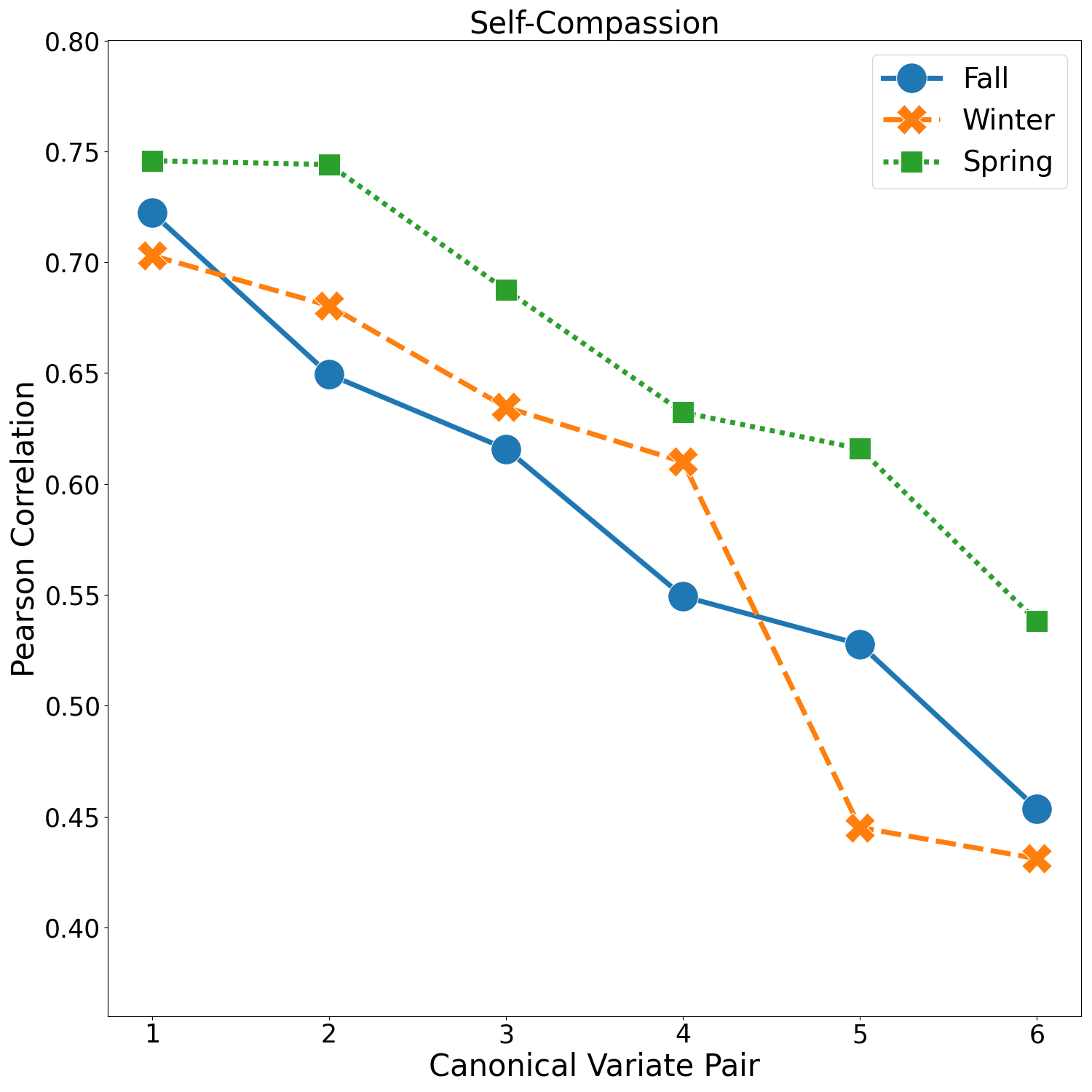}
 \includegraphics[width=0.49\linewidth]{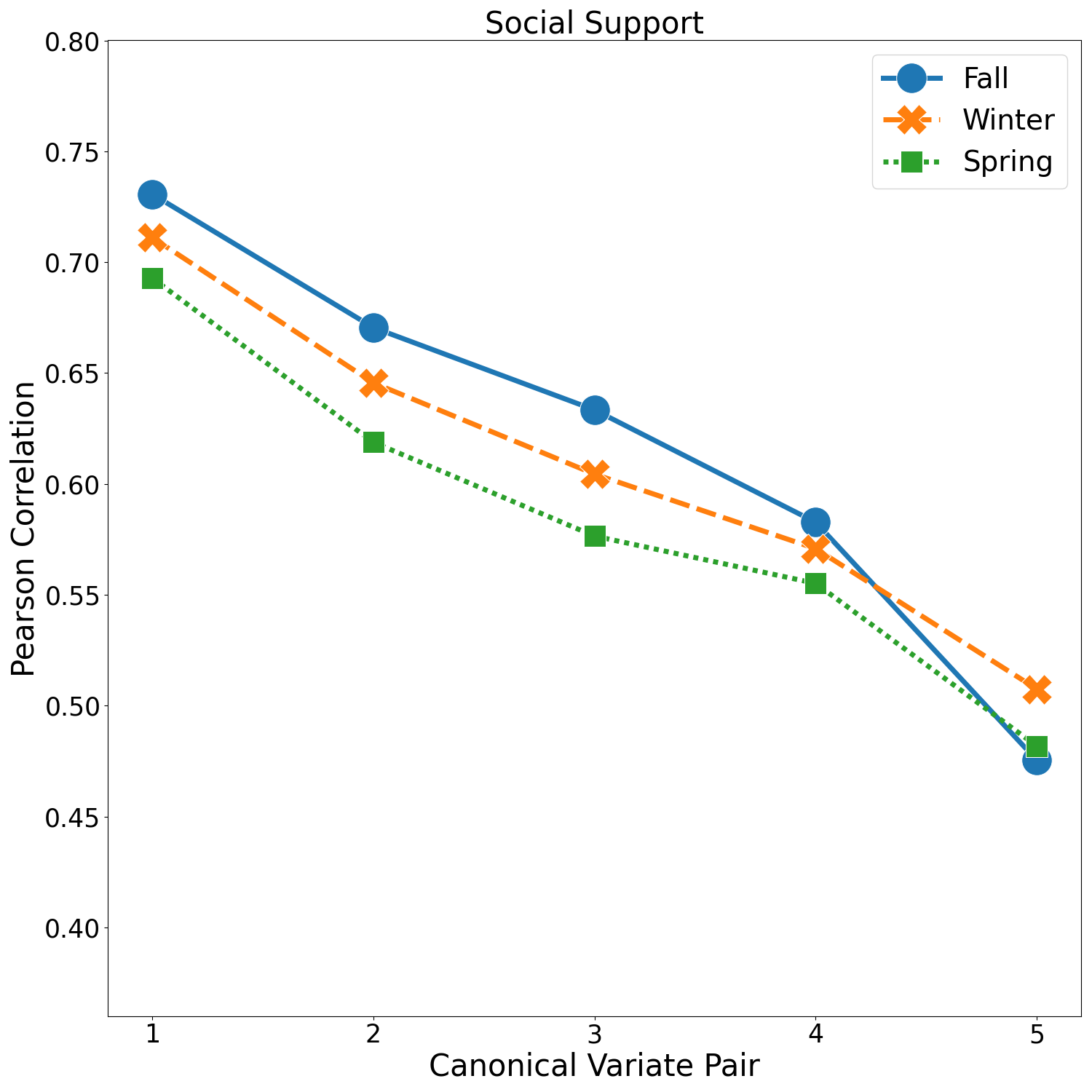}
 \caption{Fig \ref{fig:cca_change_1}a. (right) shows canonical variates coefficients corresponding to self-compassion and body composition while Fig \ref{fig:cca_change_1}b. (left) shows canonical variates for social support and body composition for the Fall, Winter, and Spring seasons. Relatively high CCA correlation was examined in the first couple of canonical variate pairs. Spring body composition and self-compassion measures had the highest correlation, while the Fall body composition and social-support measures had the highest correlation to one another.}
 \label{fig:cca_change_1}
\end{figure}

\begin{figure}[h]
 \centering
 \includegraphics[width=0.49\linewidth]{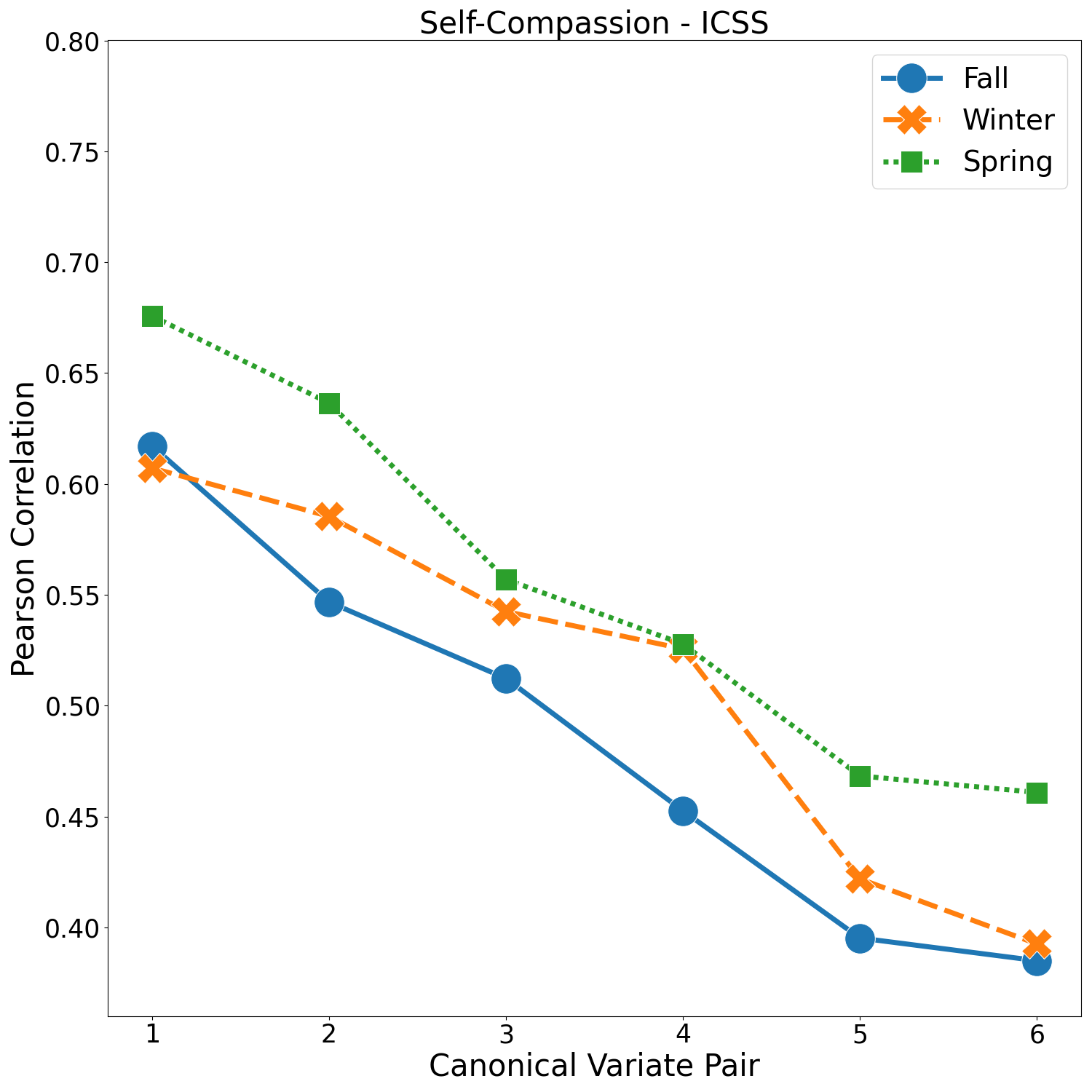}
 \includegraphics[width=0.49\linewidth]{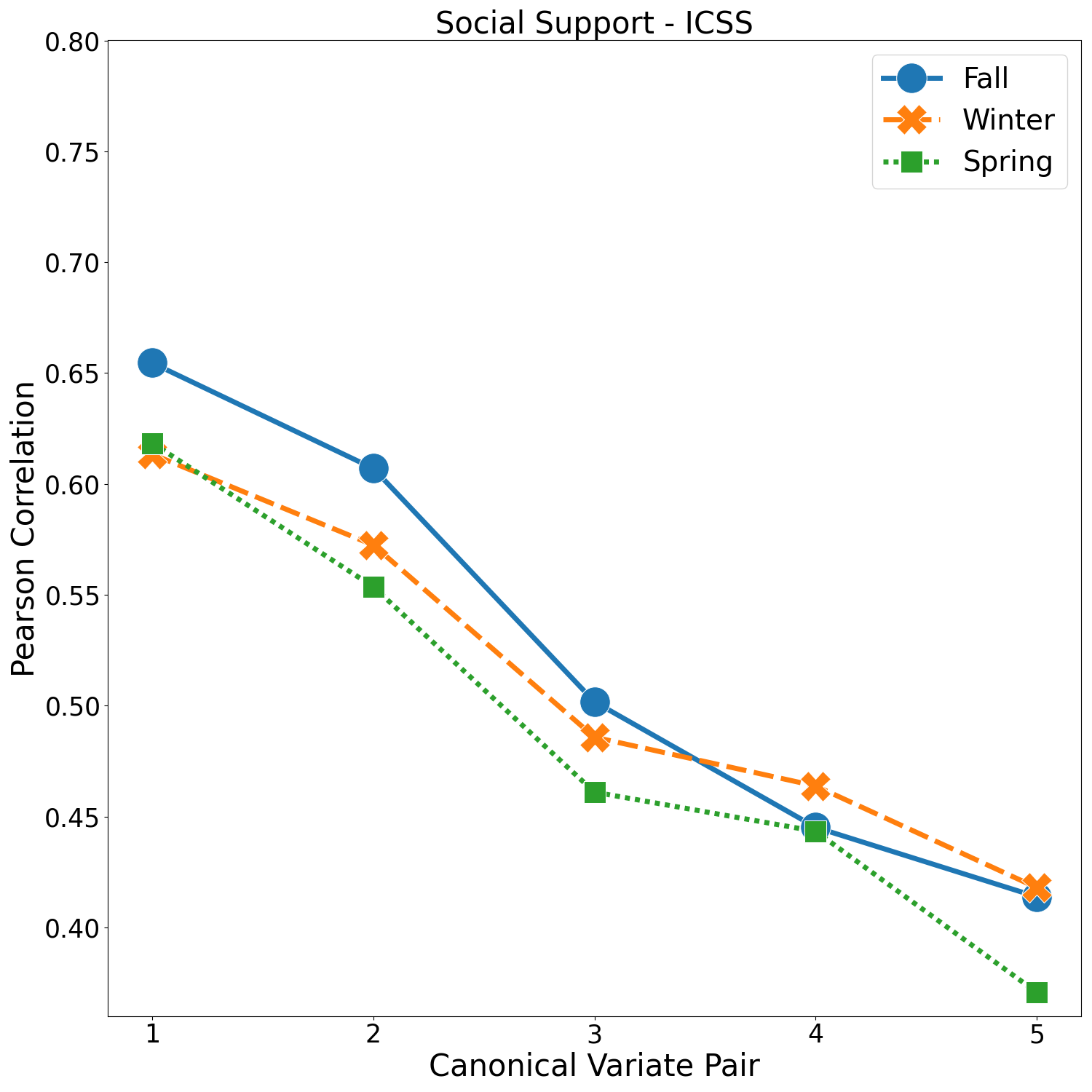}
 \caption{Fig \ref{fig:cca_change_2}a. (right) shows canonical variates coefficients corresponding to self-compassion and ICSS body composition, while Fig \ref{fig:cca_change_2}b. (left) shows canonical variates for social support and ICSS body composition for the Fall, Winter, and Spring seasons. Similar patterns were observed within the ICSS features as they were with all body composition measures.}
 \label{fig:cca_change_2}
\end{figure}

\subsubsection{Justification and Usage of Canonical Correlation Analysis}

Canonical Correlation Analysis (CCA) is a method that takes two sets of variables and obtains the linear combinations of each set that give the highest correlation. Each of these combinations is called a canonical variate, and CCA ranks pairs of canonical variates by their Pearson correlation coefficient. CCA enables the examination of relationships between two sets of variables, uncovering multivariate patterns that may not be detectable through traditional pairwise correlation approaches. We observe the Pearson correlation between the canonical variates, the latent variables from CCA that describe the original variables.

Canonical correlation analysis (CCA) is a useful method for examining relationships between two groups of variables because it identifies linear combinations of variables in each group that are maximally correlated. In doing so, CCA can provide a lower-dimensional representation of the association structure between the two sets. This makes it well-suited to our problem, where the primary goal is to quantify relationships between predefined variable groups rather than to build a predictive model. Given the limited size and diversity of the dataset, more complex machine learning pipelines may be difficult to justify and could be prone to overfitting (\cite{li_suboptimal_2024}). Additionally, standard regression-based approaches may be less appropriate for capturing multivariate relationships between the two variable sets. We therefore favor CCA over these standard methodologies.

Although a common sample-size heuristic for CCA recommends $n \geq 10 \times number \ of \ variables$, we opted against a PCA-CCA pipeline on the grounds of interpretability. Reducing body composition variables to abstract principal components would preclude meaningful inference about the original measures, sacrificing substantive interpretability for statistical convenience.

In this study, we examine group-wise correlations between InBody770 body composition metrics and two psychosocial factors: social support and self-compassion. Our task is to identify broader physiological patterns by determining which combination of variables is related to each other. 

We focus on the first two pairs of canonical covariates—$\vec{u}_1$ and $\vec{v}_1$, and $\vec{u}_2$ and $\vec{v}_2$—which exhibit the highest correlations. Here, $\vec{u}$ represents the canonical variates derived from the InBody770 body composition measurements, while $\vec{v}$ corresponds to the canonical variates derived from either the self-compassion or social support measures. To interpret these canonical variates, we examine their correlations with the original variables (denoted as X for InBody770 metrics and Y for psychosocial measures). Specifically, the correlations between $\vec{u}$ and X identify which raw InBody770 measurements most strongly contribute to $\vec{u}_1$ and $\vec{u}_2$. Likewise, the correlations between $\vec{v}$ and Y highlight the self-compassion or social support items most associated with $\vec{v}_1$ and $\vec{v}_2$. By identifying the raw variables that most strongly define each canonical variate, we can better characterize the multivariate relationships between body composition and self-compassion or social support. Correlations were labeled as significant if the p-value was less than 0.05.

To examine associations between raw measures and canonical variates, a third combination of InBody770 measures, called the InBody770 composite and segmental subset (ICSS), was examined. ICSS is comprised of composite measures, segmental impedance at 5 kHz, 50 kHz, and 500 kHz (right arm, left arm, trunk, right leg, left leg), the extracellular water to total body water ratio (ECW/TBW), systolic and diastolic blood pressure, heart rate, the InBody770 composite score, fat and lean body mass control recommendations, and segmental lean and fat mass percentages. All statistical tests were performed using the Python package sklearn (\cite{pedregosa_scikit-learn_2018}).

Finally, the sub-scales used in our study employ the same psychometric analysis protocol as described in each original study (see Background). To assess the prevalence of seasonal changes, we compared shifts in body composition, self-compassion, and social support between freshmen and sophomores across the fall, winter, and spring semesters. After splitting the dataset into freshmen and sophomores, we graphed the average value of each psychoanalysis composite measure (PACM) over the year. We also analyzed the differences between seasonal PACMs via violin plots. 

We conducted two one-way ANOVA (\cite{fisher_statistical_1925}) analyses. First, to examine whether self-compassion, social support, and InBody770 measures differed significantly across three groups defined by academic term: fall, winter, and spring. Second, we examined whether self-compassion and social support differed significantly between freshmen and sophomores for each season. The ANOVA tested for overall differences in mean self-compassion, social support, and InBody770 scores among these groups.

\section*{Results}

\subsubsection{CCA Results}
When examining CCA (Fig \ref{fig:cca_change_1}), self-compassion and social support, PACMs were strongly correlated with InBody770 measures. Pearson correlation coefficients between the canonical variates for body composition and social support range between 0.46 and 0.73 for all seasons, while correlations between the canonical variates for body composition and self-compassion range from 0.43 and 0.75 (Fig \ref{fig:cca_change_1}). In contrast, PACMs were not strongly correlated with solely the composite InBody770 measures. Pearson correlations between canonical variates for composite InBody770 measures and social support range between 0.14 and 0.44 for all seasons, while correlations between canonical variates for composite InBody770 measures and self-compassion range from 0.09 and 0.42 for all seasons. PACMs were strongly correlated with the ICSS (Fig \ref{fig:cca_change_2}). Pearson correlations between canonical variates for ICSS and social support range between 0.38 and 0.67 for all seasons, while correlations between canonical variates for ICSS and self-compassion range from 0.37 and 0.65. Canonical correlations analysis demonstrated that there are strong, significant correlations (absolute value range: 0.317 - 0.888) between features and variates for certain seasons (Table \ref{tab:cca_moss_ux}, \ref{tab:cca_selfcompassion_ux}, \ref{tab:cca_moss_vy}, \& \ref{tab:cca_selfcompassion_vy}).

\subsubsection{ANOVA Results}

ANOVA results yield no significant difference between seasonal measures of PACMs and composite InBody770 measures (Table \ref{table:anova_fvsS_fall}, \ref{table:anova_fvsS_spring},\ref{table:anova_fvsS_winter}), and freshmen vs sophomores (Table \ref{table:anova_pacms}, \ref{table:anova_inbody}). Means were also reported to show how the average differences fell among these groupings (Table \ref{table:seasons_avg}, \ref{table:freshmenvssophomore_avg}). Bonferroni or false discovery rate (FDR) correction was not applied, as the unadjusted p-values were not statistically significant.

\section*{Discussion}
\subsection{InBody770 Measures Are Correlated with Self-Compassion and Social Support PACMs.}

Our CCA results demonstrate that body composition metrics are correlated with self-compassion and social support. As shown in Fig 1, CCA demonstrates a high correlation between body composition metrics and PACMs that is consistent across various seasons. These results indicate that InBody770 measures may be useful indicators of self-compassion and social support, reinforcing previous research that examines relationships between body composition, diet, and mental health (\cite{fulton_menace_2022, segal_psychological_2025}). These findings further support research that looks for alternative measures to BMI by demonstrating that InBody770 metrics offer a complementary method of examining potential proxies for psychosocial well-being, and more broadly, mental health indicators.

The InBody770 measures taken in the fall exhibited a higher correlation with social support (Fig. 1a) compared to self-compassion (Fig. 1b). In contrast, in spring, these measures showed a weaker correlation with self-compassion and a stronger relationship with social support, potentially indicating a change in the factors associated with psychological well-being over time. This seasonal variation is consistent with broader research on college student adjustment, in which social connectedness and internalized coping strategies such as self-compassion have each been associated with psychological well-being at different points in the academic year(\cite{worsley_bridging_2021,zhou_characteristics_2023,ross_distress_2025,lavin_role_2020,liu_self-compassion_2024}).

\subsection{Analyzing PACM Contributions to CCA Covariates via Seasonality.}

When examining the relationships and associations between canonical variates and the original data, we reveal how the most representative physiological patterns relate to the broader body composition profile in the context of self-compassion and social support and vice versa. Our results show low to moderate CCA correlations (0.38-0.4) when utilizing the composite InBody770 measures, which indicates that there is important information in the stratified raw measurements. We chose to examine ICSS as a way to re-implement more detailed InBody770 measurements into our dataset while removing features that provided redundant information. This yielded stronger CCA correlations ranging from 0.37 to 0.67.\\

To better understand the relationships uncovered by canonical correlation analysis, we focus on the first two pairs of canonical covariates—$\vec{u}_1, \vec{v}_1$ and $\vec{u}_2, \vec{v}_2$—which show the strongest correlations between the two variable sets. Here, $\vec{u}$ represents the canonical variates derived from the InBody770 body composition data, while $\vec{v}$ represents those derived from the psychosocial measures of self-compassion or social support. By examining how these canonical variates correlate with their original variables (X for InBody770 metrics and Y for psychosocial measures), we identify which raw features most strongly contribute to each canonical dimension. This approach allows us to interpret and characterize the complex multivariate associations between physical body composition and psychosocial well-being. Statistical significance for these correlations was determined at a p-value threshold of 0.05.

\subsection{PACM Seasonal Changes Demonstrate No Differences Between Freshmen and Sophomores in LEMURS.}

Freshmen and sophomores are often assumed to have drastically different experiences of self-compassion and social support, with the expectation that as students progress beyond their first year, they develop greater emotional regulation and resilience, potentially due to increased peer-to-peer interaction (\cite{Landa2019,bloomfield_predicting_2024}). The transition to college represents a significant period of adjustment during which students must navigate increased independence, academic pressures, and the absence of pre-existing support networks. Homesickness and anxiety have been associated with increased stress, loneliness, and decreased well-being (\cite{Landa2019,bloomfield_predicting_2024}). As the academic year progresses, freshmen experience changes in their coping strategies, relying more on peer support and institutional resources to mitigate these challenges (\cite{Landa2019,bloomfield_predicting_2024}). 

Our results indicate that within this cohort, sophomores experience no significant (Table \ref{table:anova_fvsS_fall}, \ref{table:anova_fvsS_spring}, \& \ref{table:anova_fvsS_winter}) difference in perceived social support or self-compassion compared to first-year students. Furthermore, our results also indicate that there is no significant change between seasons (Table \ref{table:anova_pacms} \& \ref{table:anova_inbody}) within our cohort.

These findings may be interpreted in two ways. First, despite having or developing a stronger support network, students beyond their first year do not report higher levels of perceived social support or self-compassion. This aligns with research indicating that providing support, rather than receiving it, may be more beneficial to mental health (\cite{Xin2025}). Second, these findings are consistent with the "Sophomore Slump" phenomenon, whereby students experience decreased levels of interest, poor academic performance, and lower GPA (\cite{Hodgkin2025,Zuckerman2026}), highlighting the need to more closely examine the sophomore year as a critical period when mental health support may be vital. Although group means are reported for transparency (Table \ref{table:seasons_avg} \& \ref{table:freshmenvssophomore_avg}), no statistically significant differences were observed in the ANOVA results; therefore, post hoc comparisons and detailed interpretation of mean differences were not pursued.

Future research needs to investigate the relationship between perceived and actual levels of self-compassion and social support. Additionally, the assumed growth in stability, self-compassion, and social support that students experience as they progress through college may not align with reality. Further research is needed to understand whether students are forming the types of social networks and mental frameworks that enable them to fully benefit from increased social support and the development of self-compassion.

\subsection{CCA Canonical Variate Analysis: Common Themes}

Throughout the year, affectionate support and tangible support consistently correlated with the canonical variates, underscoring students’ need for emotional connection and practical assistance. These forms of support are especially relevant in the college context, where transitioning to a new environment often demands both emotional reassurance and help with daily tasks (\cite{Landa2019}).

Across all seasons, mindfulness and over-identification emerged as consistent contributors to body composition. This contribution aligns with research linking mindfulness to healthier daily behaviors, such as regular physical activity, balanced eating, and adequate sleep, in the context of managing negative or painful emotions. Mindfulness is also linked to over-identification (\cite{raes_construction_2011}) and body fat composition  (\cite{Preissner2026}) . The consistency of this association across seasons further corroborates these established relationships.

To date, there is no well-established evidence demonstrating the direct relationships mentioned in this section. While emerging or less widely known research may explore these connections, clear empirical support remains sparse. 

\paragraph{Social Support - Fall Comparisons}
 
When examining the top two canonical variates ($\vec{u}_1$, $\vec{u}_2$) of the U–X correlations (Table \ref{tab:cca_moss_ux})—which represent the associations between the canonical variates derived from the ICSS InBody770 measures and the original body composition variables—trunk fat mass percentage and body weight were the strongest contributors to the social support $\vec{u}_1$ in the Fall. Moving into the V-Y correlations (Table \ref{tab:cca_moss_vy}), which represent the associations between the canonical variates derived from the PACMs and the original PACMs, tangible support was the strongest contributor to the social support $\vec{v}_1$ in the Fall.

These findings are consistent with recent research showing that social support protects against abdominal fat accumulation and thus potentially reduces the risk of developing coronary heart disease later in life (\cite{wang_influence_2014}). Furthermore, trunk fat mass is a major predictor of increased Alanine Aminotransferase, an enzyme which, if elevated, is linked to liver damage (\cite{Boncan2024}).

Because $\vec{u}_1$ and $\vec{v}_1$ are the most highly correlated linear combinations of InBody770 measurements and social support measures, respectively, we can say that trunk fat mass percentage and body weight are correlated with tangible support. Some research has shown there to be importance to children's weight when studying parental support for physical activity (\cite{brunet_perceived_2014}). However, there is limited to no well-established evidence that suggests that there is a biological relationship between trunk fat mass and tangible support. 

Unlike the first canonical variate corresponding to the InBody770 Measures ($\vec{u}_1$), which were most correlated with weight and trunk fat mass, the second canonical variate ($\vec{u}_2$) could not be summarized by a few of the original InBody770 measurements. These indicate that the contributions for the second canonical variate for body composition are more evenly distributed across the body composition metrics. On the other hand, the corresponding second canonical variate for social support, $\vec{v}_2$, is dominated by emotional support. Thus, we can say that emotional support is related to body composition in general, although we cannot pinpoint which InBody770 metrics.\\

\paragraph{Social Support - Winter Comparisons}

Transitioning into the Winter, electrical impedance measured at 500 kHz in the trunk region and ratio of extracellular water (ECW) to total body water (TBW) emerged as significant contributors to social support $\vec{u}_1$; while emotional support and affectionate support emerged as the top contributors to social support $\vec{v}_1$. 

There is little to no evidence showing relationships between trunk impedance, ECW, emotional support, and/or affectionate support. However, on their own, they are important biomarkers for serious physical and mental health issues. For example, research has shown in female rowers that higher ECW, higher self-acceptance, and better physical condition are actually attributed to more negative weight-related pressures (the internal or external stress a person feels to control, reduce, or maintain a certain body weight) (\cite{larrinaga_eating_2024}).

Our results further link ECW/TBW, in combination with trunk electrical impedance, as an important contributor to emotional and affectionate support. Our results further implicate ECW/TBW, in combination with trunk-specific electrical impedance at 500 kHz, as a key contributor to social support $\vec{u}_1$ and $\vec{u}_2$. Furthermore, ECW/TBW is a validated marker for cancer (\cite{horino_extracellular_2023}) and Chronic Obstructive Pulmonary Disease (\cite{xie_extracellular_2024}), while elevated ECW/TBW is linked to poorer physical functioning (\cite{ishiyama_extracellular--total_2024}). If validated, these associations could show how shifts in hydration and cellular integrity relate to the etiology of psychological disorders and chronic or life-threatening diseases.

For $\vec{u}_2$, the electrical impedance measured at 5, 50, and 500kHz of the right leg was a significant contributor, however, with low correlative power (0.31 to 0.33); while positive social interaction and affectionate support were the top contributors to social support $\vec{v}_2$. 

Similarly, there is little to no evidence showing relationships between leg impedance, positive social interaction, and/or affectionate support. However, when examining leg impedance alone, these findings are consistent with research indicating that calf measurements can serve as significant markers for overall health (\cite{scott_calf_2018}).\\

\paragraph{Social Support - Spring Comparisons}

During the Spring, no strong U-X correlations emerged or persisted in the social support $\vec{u}_1$, $\vec{u}_2$, and $\vec{v}_1$, which demonstrates that for social support in the Spring, contributions ICSS variables were more evenly distributed and vice versa. 

Positive social interaction and tangible support were the most contributing variables to social support $\vec{v}_2$. Thus, we can say that positive social interaction and tangible support are related to body composition in general, although we cannot pinpoint which InBody770 metrics.
\\

\paragraph{Self-Compassion - Fall Comparisons}

Examining the self-compassion U–X correlations (Table \ref{tab:cca_selfcompassion_ux}), no significant U-X correlations emerged in the Fall; while isolation and mindfulness were significant contributors to the Fall self-compassion $\vec{v}_1$. For $\vec{v}_2$, self-judgment and over-identification were significant contributors in the Fall. 

While isolation, mindfulness, self-judgment, and over-identification were dominant contributors to $\vec{v}_1$ and $\vec{v}_2$, these measures were related to a wide range of InBody770 measures. Considering the importance of Fall as a transitional period between Summer break and the school year, it is possible that, like social support in the Spring, no single physiological metric dominates.

First, the transition to college represents a significant period of adjustment, during which students must navigate increased independence, academic pressures, and the absence of preexisting support networks. Homesickness and anxiety have been linked to heightened stress, loneliness, and decreased well-being (\cite{Landa2019, bloomfield_anxietypredictors_2024}), which may contribute to fluctuations in mindfulness and self-compassion-related measures.

\paragraph{Self-Compassion - Winter Comparisons}

Transitioning into Winter, there were no significant contributors to $\vec{u}_1$, while isolation and common humanity were the strongest contributors to self-compassion $\vec{v}_1$. There is no well-established evidence demonstrating a connection between isolation and common humanity and body composition metrics. 

The ratio of ECW to TBW and the percentage of lean body mass (LBM) in the trunk region emerged as significant contributors to the self-compassion $\vec{u}_2$; while and self-kindness and mindfulness were the strongest contributors to self-compassion $\vec{v}_2$. 

Again, there is no well-established evidence demonstrating a linkage between ECW to TBW and/or self-kindness. However, there is strong evidence supporting a connection between mindfulness and body fat, particularly in the trunk region (\cite{Preissner2026}). 

\paragraph{Self-Compassion - Spring Comparisons}

Lastly, in the Spring, mindfulness became the strongest contributor to self-compassion $\vec{v}_1$, while self-judgment and over-identification were the strongest contributors to self-compassion $\vec{v}_2$.

\subsection{Strengths and Weaknesses}

The present study found significant associations between body composition measures and self-reported social support and self-compassion; however, these findings do not support causal claims. Three problems remain: 1) whether body composition influences perceived social support and self-compassion; 2) whether those psychosocial variables shape body composition outcomes; and 3) whether unmeasured third variables account for the observed associations. Longitudinal or experimental designs are needed to establish the direction and mechanism of these relationships.

Due to the limitations of the LEMURS data, data on junior and senior class years were not available to test whether self-reported self-compassion and social support measures continued to strengthen or degrade.  As a result of our data limitations, the current study does not accurately reflect the entire population from which the LEMURS data is derived. 

Without taking into consideration major life events centered around data collection, it is impossible to have a dataset that is unbiased towards students’ lives and circumstances. For example, the data collected in January coincided with students returning from winter break, a time that may reflect elevated well-being due to recent rest and time at home. In contrast, data collected in April occurred just before finals, a period typically associated with heightened academic stress. These temporal factors may have influenced self-reported measures, introducing potential bias based on the academic calendar rather than stable psychological states. (Note: these biases could also be interpreted as important due to the study's focus on college students.) We also acknowledge that this study is based on a tiny sample from one of thousands of colleges/universities and carries inherent biases.

A further limitation of our study is the reliance on self-reported data for self-compassion and social support. Self-report is subject to recall bias and social desirability bias (\cite{Teh2023}); sensitive constructs such as psychological distress or mental health service use are particularly vulnerable, as respondents may under-report due to stigma. Self-reported body composition and health behaviors may also diverge from objectively measured values, and the resulting measurement error could attenuate or inflate observed associations. Because all constructs were assessed within the same surveys, common method variance may have contributed to the observed relationships independent of any true underlying association. Future research using objective measures (e.g., clinician-assessed body composition, validated diagnostic interviews) or multi-informant data would help determine whether the patterns reported here reflect genuine relationships or shared measurement artifacts.

Gaining a deeper understanding of these temporal factors and their impact on psychological states is essential for determining when additional resources should be allocated to support students’ needs. Moreover, since each student responds differently to stressors, understanding individual reactions to these temporal patterns could inform the development of more personalized and effective interventions.

Another notable limitation of this study is that it does not examine the nuanced and multifaceted influences on a student’s self-image and peer comparison, such as cultural context, socioeconomic status, clothing choices, and racial identity. While body composition or BMI may contribute to self-perception, these additional confounding variables likely play a significant role in shaping how students view themselves and others. 

Finally, our study is phenotype-blind, meaning we do not account for individual biological or physical characteristics in our analysis, such as hydration, activity, stress cycles, race, biological sex, sexual activity, and cultural epigenetics. As a result, any population-level trends we observe may obscure meaningful subgroup differences and reduce the specificity of our findings.

The absence of these factors from the analysis limits the study’s ability to fully capture the complexity of the correlation, and a longitudinal design may be better suited to explore these dynamics over time.

\section*{Conclusion}

Our results demonstrate a clear correlation between body composition measures and self-perceived psychosocial measures. Self-compassion and social support measures did not differ between freshmen and sophomores. Further analysis is needed to find the root cause of these non-differences.  

While no class-year differences were observed in self-compassion or social support, consistent associations emerged between well-being and specific physiological and psychosocial factors—most notably, trunk and leg impedance, mindfulness, over-identification, affectionate support, and tangible support. Until now, these associations have had limited empirical support.

These findings have practical implications for students, campus mental health services, and wellness programs. Body composition measures are non-invasive and low-cost relative to standard psychological assessments, making them an accessible complement to existing screening approaches. If these associations are replicated, colleges could incorporate body composition metrics into routine health assessments to identify students whose profiles are associated with lower self-compassion or reduced perceived social support. Those students could then be connected with targeted resources, including counseling services, peer support networks, and campus wellness programs focused on nutrition, physical activity, and stress management. 

Proactively identifying at-risk students could improve individual well-being and broader institutional outcomes, because research consistently links poor mental health and low social connectedness to academic disengagement and attrition (\cite{chan_impact_2022, wang_association_2024, wei_first-year_2022}). This application may be especially relevant given the persistent gap between mental health need and service use in college populations. Studies estimate that a substantial proportion of college students report clinically meaningful distress but do not access campus mental health resources, with under-utilization attributed to stigma, limited awareness, and financial barriers (\cite{aluri_prevalence_2025, rackoff_psychotherapy_2025, ncha_survey_summary_2024}). Embedding a mental health-relevant indicator within a non-stigmatized health assessment may reduce barriers to identification and referral, supporting both student well-being and institutional retention goals.

Future work could further explore the causal relationships between seasonal changes in psychosocial and physiological well-being, between self-compassion and social support as measured in freshmen and sophomores.

Other types of CCA (ie. probabilistic, ridge, or regularized CCA) could provide more robust inference than regular CCA due to being more applicable to our small sample size and high dimensionality (\cite{yang_survey_2021}).

An examination of a PCA-CCA pipeline could provide insight into how to align small biological datasets with standard common sample-size heuristics and how to extract interpretable results from linear combinations of biological data.

\section*{Disclosure Statement}

The authors have no conflicts of interest to report. The authors confirm that the research presented in this article met the ethical guidelines, including adherence to the legal requirements, of the United States and received approval from the Institutional Review Board of the University of Vermont.


\section*{Acknowledgment}
This work was supported in part by a grant from MassMutual and the Summer Undergraduate Research Fellowship (SURF) at the University of Vermont.

\printbibliography



\begin{appendix}

\subsection{"Raw" or Body Composition Measures} The entire feature set derived from the InBody770 machine.

\subsection{Composite Body Composition Measures} Measures relating to entire feature sets. These include Weight (lbs), Total Body Water, Intracellular Water, Extracellular Water, Dry Lean Mass, Body Fat Mass, Lean Body Mass, Skeletal Muscle Mass, and Percent Body Fat.

\subsection{ICSS} A subset of the entire InBody770 feature set, including the \textit{composite body composition measures} and segmental impedance at 5 kHz, 50 kHz, and 500 kHz (right arm, left arm, trunk, right leg, left leg), the extracellular water to total body water ratio (ECW/TBW), systolic and diastolic blood pressure, heart rate, the InBody770 composite score, fat and lean body mass control recommendations, and segmental lean and fat mass percentages.

\end{appendix}


\clearpage
\section*{Supplementary Material}
\setcounter{table}{0}
\setcounter{figure}{0}
\renewcommand{\thetable}{S\arabic{table}}
\renewcommand{\thefigure}{S\arabic{figure}}

\begin{table}[h]
\centering

\caption{Canonical Correlation Coefficients ($\rho$) for ICSS. First reported are the canonical variates coefficients corresponding to self-compassion and ICSS body composition. Second reported are the canonical variates coefficients for social support and ICSS body composition. Note the decreased correlation power, clearly noting that the removed features (segmentation of composite body composition variables) added information. This may suggest that analyzing body composition features by specific body regions could provide greater insight into psychosocial measures.}
\begin{tabular}{|lcccccc|}
\hline

Season & $\rho_1$ & $\rho_2$ & $\rho_3$ & $\rho_4$ & $\rho_5$ & $\rho_6$ \\
\hline

& & & & & &\\
\multicolumn{7}{|c|}{\textbf{Self-Compassion}} \\
Fall& 0.6171 & 0.5467 & 0.5122 & 0.4526 & 0.3952 & 0.3850 \\
Winter & 0.6071 & 0.5854 & 0.5425 & 0.5256 & 0.4219 & 0.3926 \\
Spring & 0.6757 & 0.6365 & 0.5570 & 0.5278 & 0.4683 & 0.4609 \\
& & & & & &\\

\multicolumn{7}{|c|}{\textbf{Social Support}} \\
Fall& 0.6546 & 0.6070 & 0.5017 & 0.4451 & 0.4138 & \\
Winter & 0.6134 & 0.5724 & 0.4858 & 0.4638 & 0.4179 & \\
Spring & 0.6184 & 0.5536 & 0.4610 & 0.4433 & 0.3707 & \\
\hline
\end{tabular}
\label{tab:CCA-Values-ICSS}
\end{table}

\begin{table}[h]
\centering
\caption{Canonical Correlation Coefficients ($\rho$) summarizing the strength of association between the body composition variables and self-compassion/social support. First reported are the canonical variates coefficients corresponding to self-compassion and body composition. Second reported are the canonical variates coefficients for social support and body composition. Note the high correlation between self-compassion and social support to body composition.}
\begin{tabular}{|lcccccc|}
\hline

Season & $\rho_1$ & $\rho_2$ & $\rho_3$ & $\rho_4$ & $\rho_5$ & $\rho_6$ \\
\hline
& & & & & & \\
\multicolumn{7}{|c|}{\textbf{Self Compassion}} \\
Fall& 0.7222 & 0.6494 & 0.6158 & 0.5493 & 0.5277 & 0.4536 \\
Winter & 0.7030 & 0.6804 & 0.6345 & 0.6100 & 0.4450 & 0.4310 \\
Spring & 0.7458 & 0.7440 & 0.6875 & 0.6323 & 0.6158 & 0.5381 \\
& & & & & & \\

\multicolumn{7}{|c|}{\textbf{Social Support}} \\
Fall& 0.7307 & 0.6705 & 0.6334 & 0.5827 & 0.4753 & \\
Winter & 0.7112 & 0.6456 & 0.6044 & 0.5708 & 0.5075 & \\
Spring & 0.6928 & 0.6188 & 0.5765 & 0.5551 & 0.4818 & \\
\hline
\end{tabular}
\label{tab:CCA-Values-TOTAL}
\end{table}

\begin{table*}[t]
\centering
\caption{$\vec{u}$ Corr X - Canonical Correlations Between MOSS InBody770 Features and MOSS InBody770 Canonical Variates}
\vspace{8pt}
\begin{tabular}{|l|l|c|c||l|c|c|}
\hline
\textbf{Season} & \textbf{InBody770 MOSS $\vec{u}_1$} & \textbf{Corr.} & \textbf{p-value} & \textbf{InBody770 MOSS $\vec{u}_2$} & \textbf{Corr.} & \textbf{p-value} \\
\hline
\multirow{3}{*}{Fall} 
& Weight & -0.400 & $<$0.05 &  &  &  \\
& Fat Mass Adj. & -0.352 & $<$0.05 &  &  &  \\
& \% Fat Mass (Trunk) & 0.335 & $<$0.05 &  &  &  \\
\hline
\multirow{3}{*}{Winter} 
& Impedance 500kHz (Trunk) & 0.423 &  & Impedance 5kHz (Right Leg) & 0.496 &  \\
& ECW/TBW Ratio & 0.418 &  & Impedance 50kHz (Right Leg) & 0.449 &  \\
& Impedance 50kHz (Trunk) & 0.402 &  & Impedance 500kHz (Right Leg) & 0.421 &  \\
\hline
\multirow{1}{*}{Spring} 
&  &  &  &  &  &  \\
\hline
\end{tabular}
\label{tab:cca_moss_ux}
\end{table*}

\begin{table*}[t]
\centering
\caption{$\vec{u}$ Corr X - Canonical Correlations Between Self-Compassion InBody770 Features and Self-Compassion InBody770 Canonical Variates}
\vspace{8pt}
\begin{tabular}{|l|l|c|c||l|c|c|}
\hline
\textbf{Season} & \textbf{InBody770 Self Compassion $\vec{u}_1$} & \textbf{Corr.} & \textbf{p-value} & \textbf{InBody770 Self Compassion $\vec{u}_2$} & \textbf{Corr.} & \textbf{p-value} \\
\hline
\multirow{1}{*}{Fall} 
&  &  &  &  &  &  \\
\hline
\multirow{2}{*}{Winter} 
&  &  &  & ECW/TBW Ratio & -0.413 & $<$0.05 \\
&  &  &  & \% Lean Mass (Trunk) & 0.397 & \\
\hline
\multirow{3}{*}{Spring} 
& Impedance 5kHz (Right Leg) & 0.332 &  &  &  &  \\
& Impedance 50kHz (Right Leg) & 0.330 &  & &  & \\
& Impedance 500kHz (Right Leg) & 0.317 &  &  &  &  \\
\hline
\end{tabular}
\label{tab:cca_selfcompassion_ux}
\end{table*}

\begin{table*}[t]
\centering
\caption{$\vec{v}$ Corr Y - Canonical Correlations Between MOSS Features and MOSS Canonical Variates}
\begin{tabular}{|l|l|c|c||l|c|c|}
\hline
\textbf{Season} & \textbf{MOSS $\vec{v}_1$} & \textbf{Corr.} & \textbf{p-value} & \textbf{MOSS $\vec{v}_2$} & \textbf{Corr.} & \textbf{p-value} \\
\hline
\multirow{1}{*}{Fall} 
& Tangible Support & 0.421 & $<$0.05 & Emotional Support & 0.318 & $<$0.05 \\
\hline
\multirow{3}{*}{Winter} 
& Emotional Support & -0.714 & $<$0.05 & Positive Social Interaction & 0.837 & $<$0.05 \\
& Affectionate Support & -0.651 & $<$0.05 & Affectionate Support & 0.744 & $<$0.05 \\
& Tangible Support & -0.644 & $<$0.05 & Overall Social Support & 0.728 & $<$0.05 \\
\hline
\multirow{3}{*}{Spring} 
&  &  &  & Positive Social Interaction & 0.695 & $<$0.05 \\
&  &  &  & Tangible Support & 0.615 & $<$0.05 \\
&  &  &  & Overall Social Support & 0.568 & $<$0.05 \\
\hline
\end{tabular}
\label{tab:cca_moss_vy}
\end{table*}

\begin{table*}[t]
\centering
\caption{$\vec{v}$ Corr Y - Canonical Correlations Between Self-Compassion Features and Self-Compassion Canonical Variates}
\vspace{8pt}  
\begin{tabular}{|l|l|c|c||l|c|c|}
\hline
\textbf{Season} & \textbf{Self-Compassion $\vec{v}_1$} & \textbf{Corr.} & \textbf{p-value} & \textbf{Self-Compassion $\vec{v}_2$} & \textbf{Corr.} & \textbf{p-value} \\
\hline
\multirow{3}{*}{Fall} 
& Isolation & 0.648 & $<$0.05 & Self-Judgement & -0.888 & $<$0.05 \\
& Mindfulness & 0.314 & $<$0.05 & Over Identification & -0.815 & $<$0.05 \\
& Over Identification & 0.309 & $<$0.05 & Self-Kindness & -0.576 & $<$0.05 \\
\hline
\multirow{3}{*}{Winter} 
& Isolation & 0.756 & $<$0.05 & Self-Kindness & 0.783 & $<$0.05 \\
& Common Humanity & 0.628 & $<$0.05 & Mindfulness & 0.706 & $<$0.05 \\
& Over Identification & 0.450 & $<$0.05 & Over Identification & 0.544 & $<$0.05 \\
\hline
\multirow{2}{*}{Spring} 
& Mindfulness & 0.766 & $<$0.05 & Self-Judgement & -0.438 & $<$0.05 \\
&  &  &  & Over Identification & -0.395 & $<$0.05 \\
\hline
\end{tabular}
\label{tab:cca_selfcompassion_vy}
\end{table*}

\begin{table}[h!]
\centering
\caption{ANOVA one-way statistical analysis to determine the statistical seasonal and difference between Fall, Winter, and Spring measures of self-compassion and social support PACMS—composite measures isolated for analysis. P-values showed that there was no statistical significance between seasonal measures of self-compassion and social support. Analysis utilized pre-processed data.}
\begin{tabular}{|l|c|}
\hline
\textbf{Measure} & \textbf{ANOVA P-Value} \\
\hline
Self Kindness & 0.674 \\
Self Judgment & 0.452 \\
Common Humanity & 0.682 \\
Isolation & 0.866 \\
Mindfulness & 0.736 \\
Over Identification & 0.672 \\
\textit{Total Self Compassion} & 0.883 \\
 & \\
Emotional Support & 0.166 \\
Tangible Support & 0.206 \\
Affectionate Support & 0.132 \\
Positive Social Interaction & 0.310 \\
\textit{Overall Social Support} & 0.202 \\
\hline
\end{tabular}
\label{table:anova_pacms}
\end{table}

\begin{figure}
 \centering
 \includegraphics[width=\linewidth,height=0.85\textheight,keepaspectratio]{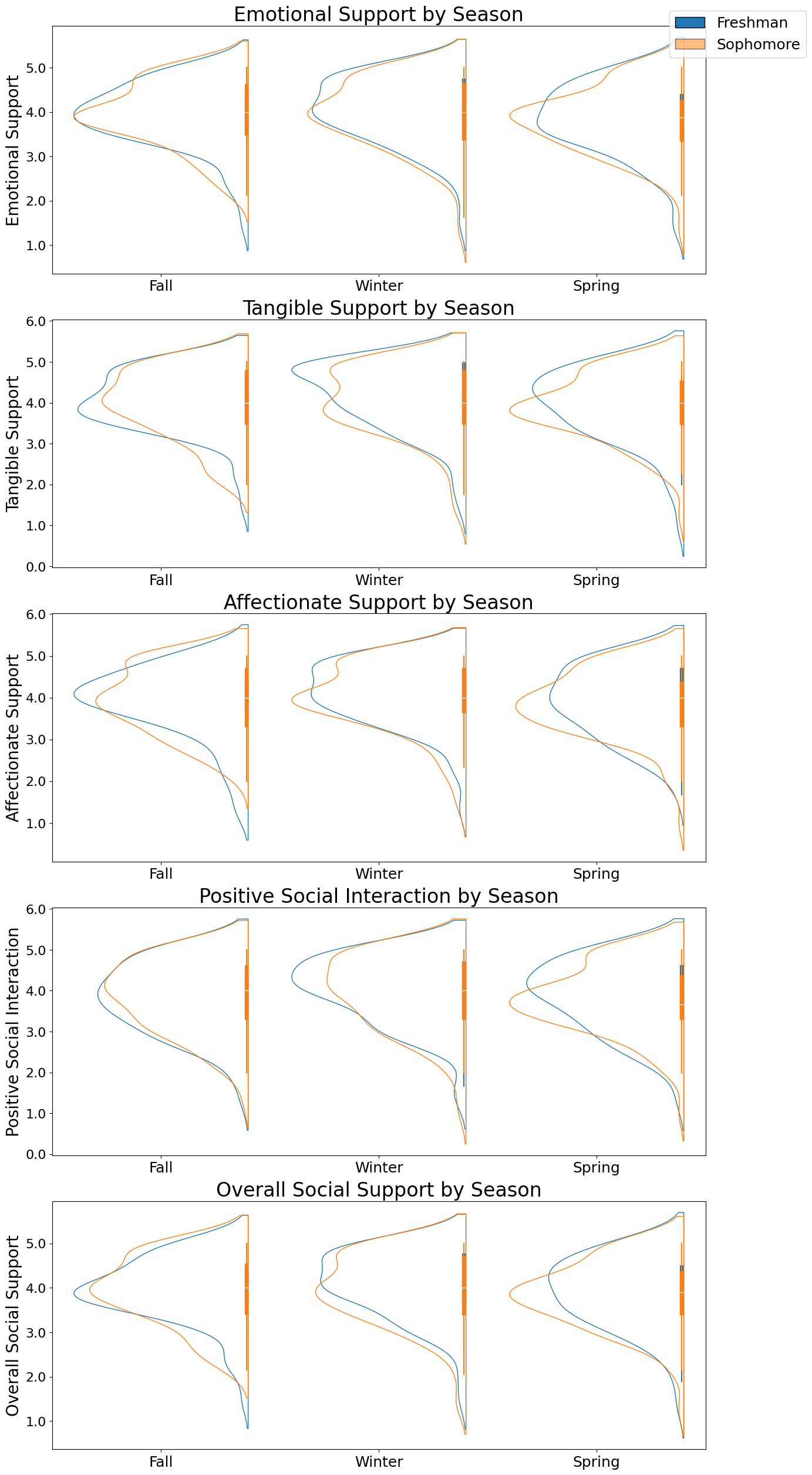}
 \caption{These violin graphs illustrate how the social support measures change across the seasons for both freshmen and sophomores. While the differences in tangible support and overall social support between freshmen and sophomores did not reach statistical significance, the effect size suggests potential practical relevance worth further exploration.}
 \label{fig:masc_moss}
\end{figure}

\begin{figure}
 \centering
 \includegraphics[width=\linewidth,height=0.85\textheight,keepaspectratio]{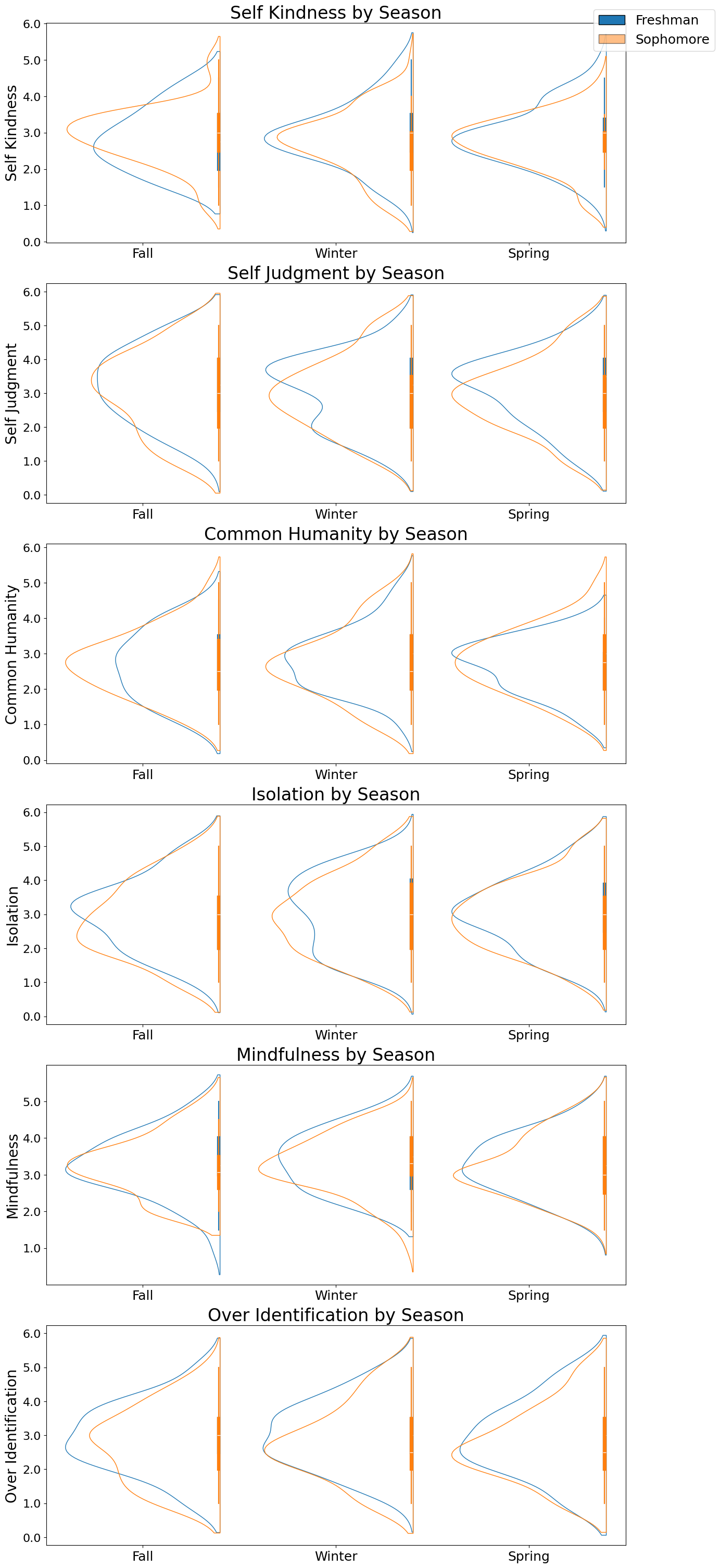}
 \caption{These violin graphs illustrate how the self-compassion measures change across the seasons for both freshmen and sophomores. While the differences in isolation and self-judgment between freshmen and sophomores did not reach statistical significance, the effect size suggests potential practical relevance worth further exploration.}
 \label{fig:masc_self-compassion}
\end{figure}

\begin{figure}[t]
 \centering
 \includegraphics[width=\linewidth,height=0.85\textheight,keepaspectratio]{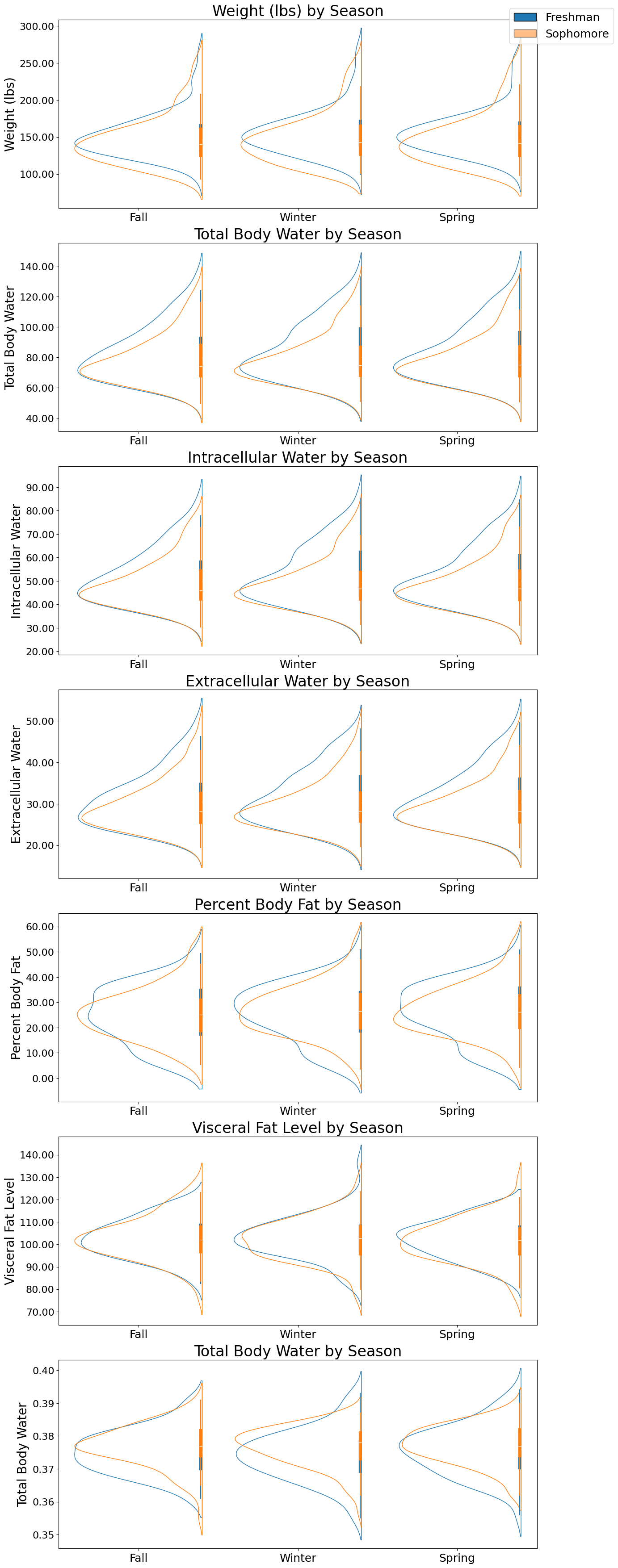}
 \caption{These violin graphs illustrate how the composite InBody770 measures change across the seasons for both freshmen and sophomores. Only selected measures of interest are shown. }
 \label{fig:masc_inbody}
\end{figure}

\begin{table}[h!]
\centering
\caption{ANOVA one-way statistical analysis to determine the statistical difference between Fall, Winter, and Spring InBody770 measures. Only selected measures of interest are shown.}
\label{table:anova_inbody}
\begin{tabular}{|l|c|}
\hline
\textbf{Measure} & \textbf{ANOVA P-Value} \\
\hline
Weight (lbs) & 0.610 \\
Total Body Water & 0.955 \\
Intracellular Water & 0.940 \\
Extracellular Water & 0.976 \\
Dry Lean Mass & 0.925 \\
Body Fat Mass & 0.485 \\
Lean Body Mass & 0.948 \\
Skeletal Muscle Mass & 0.941 \\
Body Mass Index & 0.551 \\
Percent Body Fat & 0.467 \\
\hline
\end{tabular}
\end{table}

\begin{table}[h!]
\centering
\caption{ANOVA one-way statistical analysis to determine the statistical seasonal and difference between fall freshmen and sophomore measures of self-compassion and social support PACMS—composite measures isolated for analysis.}
\begin{tabular}{|l|c|}
\hline
\textbf{Measure} & \textbf{ANOVA P-Value} \\
\hline
Self Kindness & 0.666 \\
Self Judgment & 0.226 \\
Common Humanity & 0.762 \\
Isolation & 0.313 \\
Mindfulness & 0.606 \\
Over Identification & 0.161 \\
\textit{Total Self Compassion} & 0.411 \\
 & \\
Emotional Support & 0.989 \\
Tangible Support & 0.445 \\
Affectionate Support & 0.713 \\
Positive Social Interaction & 0.853 \\
\textit{Overall Social Support} & 0.872 \\
\hline
\end{tabular}
\label{table:anova_fvsS_fall}
\end{table}

\begin{table}[h!]
\centering
\caption{ANOVA one-way statistical analysis to determine the statistical seasonal and difference between winter freshmen and sophomore measures of self-compassion and social support PACMS—composite measures isolated for analysis.}
\begin{tabular}{|l|c|}
\hline
\textbf{Measure} & \textbf{ANOVA P-Value} \\
\hline
Self Kindness & 0.140 \\
Self Judgment & 0.740 \\
Common Humanity & 0.534 \\
Isolation & 0.479 \\
Mindfulness & 0.422 \\
Over Identification & 0.137 \\
\textit{Total Self Compassion} & 0.220 \\
 & \\
Emotional Support & 0.387 \\
Tangible Support & 0.225 \\
Affectionate Support & 0.394 \\
Positive Social Interaction & 0.491 \\
\textit{Overall Social Support} & 0.315 \\
\hline
\end{tabular}
\label{table:anova_fvsS_winter}
\end{table}

\begin{table}[h!]
\centering
\caption{ANOVA one-way statistical analysis to determine the statistical seasonal and difference between spring freshmen and sophomore measures of self-compassion and social support PACMS—composite measures isolated for analysis.}
\begin{tabular}{|l|c|}
\hline
\textbf{Measure} & \textbf{ANOVA P-Value} \\
\hline
Self Kindness & 0.179 \\
Self Judgment & 0.233 \\
Common Humanity & 0.455 \\
Isolation & 0.511 \\
Mindfulness & 0.711 \\
Over Identification & 0.233 \\
\textit{Total Self Compassion} & 0.307 \\
 & \\
Emotional Support & 0.698 \\
Tangible Support & 0.612 \\
Affectionate Support & 0.713 \\
Positive Social Interaction & 0.646 \\
\textit{Overall Social Support} & 0.667 \\
\hline
\end{tabular}
\label{table:anova_fvsS_spring}
\end{table}

\begin{table}[h!]
\centering
\caption{Table showing InBody770 averages, Social Support, and Self-Compassion averages between freshmen and sophomores across all seasons. Only selected InBody measures of interest are shown.}
\begin{tabular}{|l|c|c|}
\hline
\textbf{Measure} & \textbf{Freshmen Average} & \textbf{Sophomore Average} \\
\hline
Weight (lbs) & 149.68 & 141.48 \\
Total Body Water & 80.12 & 74.62 \\
Intracellular Water & 49.60 & 46.48 \\
Extracellular Water & 30.37 & 28.23 \\
Percent Body Fat (\%) & 26.10 & 25.97 \\
Visceral Fat Level & 102.77 & 102.20 \\
ECW / TBW Ratio & 0.375 & 0.377 \\
& & \\ 
Self Kindness & 2.87 & 2.77 \\
Self Judgment & 3.13 & 2.97 \\
Common Humanity & 2.72 & 2.74 \\
Isolation & 3.03 & 2.90 \\
Mindfulness & 3.32 & 3.25 \\
Over Identification & 2.96 & 2.73 \\
\textit{Total Self Compassion} & 3.01 & 2.89 \\
& & \\ 
Emotional Support & 3.94 & 3.89 \\
Tangible Support & 4.05 & 3.94 \\
Affectionate Support & 3.96 & 3.92 \\
Positive Social Interaction & 3.89 & 3.85 \\
\textit{Overall Social Support} & 3.97 & 3.90 \\

\hline
\end{tabular}
\label{table:freshmenvssophomore_avg}
\end{table}

\begin{table}[h!]
\centering
\caption{Table showing InBody770 averages, Social Support, and Self-Compassion averages between Fall, Winter, and Spring. Only selected InBody770 measures of interest are shown.}
\begin{tabular}{|l|c|c|c|}
\hline
\textbf{Measure} & \textbf{Fall} & \textbf{Winter} & \textbf{Spring} \\
\hline
Weight (lbs) & 144.05 & 146.60 & 146.70 \\
Total Body Water (lbs) & 76.60 & 76.50 & 76.70 \\
Intracellular Water (lbs) & 47.75 & 47.50 & 47.50 \\
Extracellular Water (lbs) & 28.90 & 29.00 & 28.70 \\
Percent Body Fat (\%) & 25.20 & 26.55 & 26.25 \\
Visceral Fat Level & 102.10 & 102.80 & 102.35 \\
ECW / TBW Ratio & 0.377 & 0.376 & 0.377 \\
& & &\\
Self Kindness & 3.00 & 3.00 & 3.00 \\
Self Judgment & 3.00 & 3.00 & 3.00 \\
Common Humanity & 2.50 & 2.75 & 3.00 \\
Isolation & 3.00 & 3.00 & 3.00 \\
Mindfulness & 3.06 & 3.50 & 3.00 \\
Over Identification & 3.00 & 2.75 & 2.50 \\
\textit{Total Self Compassion} & 2.96 & 2.92 & 2.92 \\
& & &\\
Emotional Support & 4.00 & 4.00 & 3.88 \\
Tangible Support & 4.00 & 4.03 & 4.00 \\
Affectionate Support & 4.00 & 4.00 & 4.00 \\
Positive Social Interaction & 4.00 & 4.00 & 4.00 \\
\textit{Overall Social Support} & 3.97 & 4.01 & 3.89 \\
\hline
\end{tabular}
\label{table:seasons_avg}
\end{table}

\end{document}